\documentclass[12pt]{article}
\usepackage{epsfig}
\begin{document}
\begin{titlepage}
\pagestyle{empty}
\begin{flushright}
        MZ-TH/06-16\\
        September 5th., 2006
\end{flushright}
\vskip 0.7in
\begin{center}
{\large {\bf Selected Topics in Top Quark Physics}}
\end{center}
\begin{center}
 
\vskip 0.1in
{\large \bf  J.G.~K\"orner$^1$ } 
\\[.4truecm]

$^1$Institut f\"ur Physik der Johannes-Gutenberg-Universit\"at,\\
  Staudinger Weg 7, D-55099 Mainz, Germany\\[.3truecm]

\vskip 0.2in {\bf Abstract}
\end{center}
\baselineskip=18pt 
\noindent
I discuss three different selected topics in top quark physics. The first topic
concerns the next-to-next-to-leading order calculation of the hadroproduction of 
top quark pairs and the role of multiple 
polylogarithms in this calculation. I report on an ongoing 
next-to-next-to-leading order calculation of heavy quark pair production in hadron
collisions where the loop--by--loop part of the calculation is about to be completed.
Calculating the loop-by-loop part allows one to take a glimpse at the mathematical
structure of the full NNLO calculation. The loop-by-loop contributions bring in a 
new class of functions introduced only eight years ago by the Russian mathematician 
Goncharov called multiple polylogarithms. The second topic concerns a 
next-to-leading order calculation of unpolarized top quark decays which are analyzed
in cascade fashion $t \rightarrow b + W^+$ followed by 
$W^+ \rightarrow l^+ + \nu_l$. Finally, I present some next-to-leading order results 
on polarized top quark decays which are analyzed in the top quark rest system. 
\end{titlepage}

\section{Introduction}

I begin my talk with a few remarks on present top quark yields at the Tevatron
and on expected top quark yields at the LHC which will start running at the end of
2007. After a slow start in early 2001 Tevatron II started reaching design peak 
luminosities of $8.5 \cdot 10^{31} {\mbox{cm}}^{-1}{\mbox{s}}^{-1}$ in 2004. The best 
weekly performance was early 2006 with a weekly integrated luminosity of 
$25 \,{\mbox {pb}}^{-1}$.
If Tevatron II could perform at this rate it would be able to collect 
$1.3\,{\mbox{fb}}^{-1}$ in a year. There has been a three months shutdown in the 
spring of
2006 with some (electron cooling) improvements on the $\bar{p}$ beam. The hope was 
that there will be a factor
two or three improvement in luminosity after the shutdown. Such a factor would 
be dearly 
needed if one wants to reach the projected total of $8 {\mbox{fb}}^{-1}$ when the 
machine
is closed down in 2009. At the time of writing this factor
has not been realized so far after a few months of post--shutdown running although
the machine is performing quite well with continuous improvements. At a c.m. energy of $\sqrt{s}= 1.96 \,{\mbox {TeV}}$ with
$\sigma(t\bar{t})\approx 6.8\, {\mbox{pb}}$ one expects around 7000 $t\bar{t}$ pairs
at each detector (CDF and DO) for an integrated luminosity of $1\, {\mbox{fb}}^{-1}$.
Single top production occurs at about $33\%$ of the $t\bar{t}$ pair production rate
but has not been detected so far.

Much bigger samples of top quarks will be available at the LHC. Due to the higher
energy of the LHC the cross section increases by a factor of 100. Also there 
will be a ten--fold increase in luminosity at the LHC. Thus one will have
$10^{7}$ $t\bar{t}$--pairs per year, or one $t\bar{t}$--pair every four seconds
at each detector (ATLAS and CMS). In a later high luminosity run there will be
another factor of ten increase in luminosity such that one will have a 
$t\bar{t}$--pair produced every half second. Again single top production occurs
at approximately one--third the rate of $t\bar{t}$--production. Singly produced 
top quarks will be highly polarized because they are produced weakly. This 
opens the way to study angular correlations between the polarization
of the top quark and its decay products which forms the third topic of this talk.

The yield of top quark pairs at the International Linear Collider (possibly 
starting in 2015) will be \,$\approx (1 - 4) \cdot 10^5/y$ depending on the c.m.
energy $\approx 360 - 800 \,{\mbox {GeV}}$. In $e^+-e^-$--interactions a high
degree of polarization of the top (or antitop) quark can be achieved through
tuning of the beam polarization.

\section{NNLO description of heavy top quark production}

The full next-to-leading order (NLO) QCD corrections to hadroproduction of heavy 
flavors were completed as early as 1988 \cite{Dawson:1988,been}. They 
have raised the leading order (LO) estimates \cite{LO:1978} but were 
still below  the experimental results on bottom quark pair production 
(see e.g. \cite{Italians}). In a recent analysis theory moved closer to experiment 
\cite{Italians}. First experimental results on hadronic $t\bar{t}$--pair production
\cite{acosta04,abazov05} are in agreement with theoretical NLO \,QCD predictions 
\cite{cacciari04,kidonakis03} within the large theoretical and experimental error bars.

A large uncertainty in the NLO calculation results from the freedom in the 
choice of the renormalization and factorization scales. These scale uncertainties 
amount to a $\approx 10\%$ theoretical error in the NLO cross section predictions
\cite{cacciari04}.
The dependence on the factorization and renormalization scales is expected to be 
greatly reduced at next-to-next-to-leading order (NNLO). This will
reduce the theoretical uncertainty. 

\begin{figure}[t]
\center
\includegraphics[height=9cm]{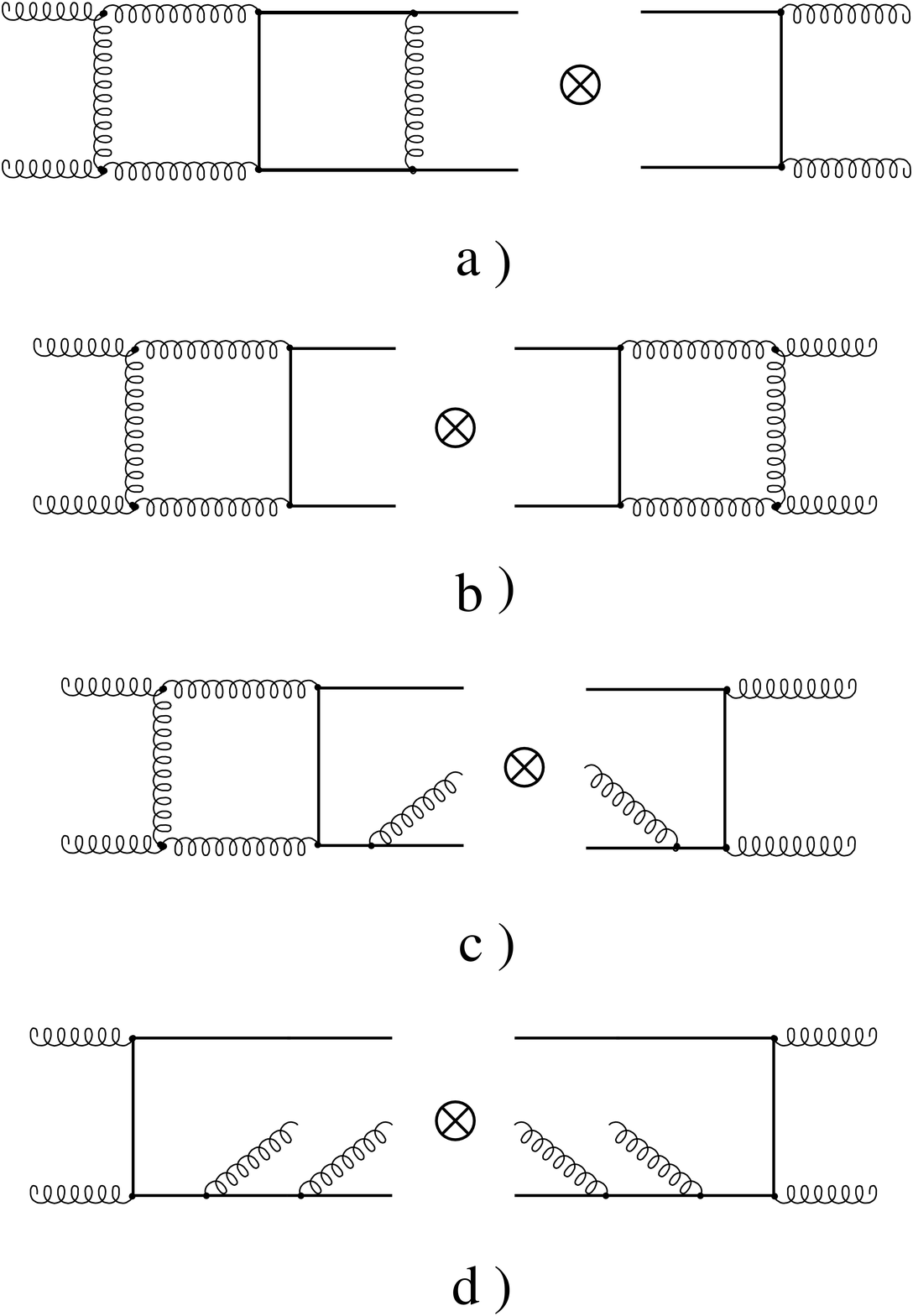}
\caption{Exemplary gluon fusion diagrams for the  NNLO calculation of heavy 
hadron production}
\label{nnlo}      
\end{figure}

In Fig.~\ref{nnlo} we show one generic diagram each for the four classes
of gluon induced contributions that need to be calculated for the NNLO corrections
to hadroproduction of heavy flavors. They involve the two-loop
contribution (Fig.~\ref{nnlo}a), the loop-by-loop contribution (Fig.~\ref{nnlo}b), 
the one-loop gluon emission contribution (Fig.~\ref{nnlo}c) and, finally, the two 
gluon emission contribution (Fig.~\ref{nnlo}d).
In our work we have concentrated on the
loop-by-loop contributions exemplified by Fig.~\ref{nnlo}b. Specifically, working 
in the framework of  dimensional regularization, we have 
calculated ${\cal O}(\epsilon^2)$ results for
all scalar massive one-loop one-, two-, three- and four-point integrals that are
needed in the calculation of hadronic heavy flavour production \cite{kmr1}. 
Because the one-loop integrals exhibit infrared (IR)/collinear (M)
singularities up to ${\cal O}(\epsilon^{-2})$ \cite{km02} one needs to know the 
one-loop 
integrals up to ${\cal O}(\epsilon^2)$ since the one-loop contributions appear 
in product form in the loop-by-loop contributions. It is exactly the 
${\cal O}(\epsilon^2)$ terms in the scalar massive three- and four-point integrals
that bring in multiple polylogarithms \cite{kmr1,kmr06a}.  

Calculating the loop-by-loop contributions allows one to obtain a glimpse of the 
complexity that is waiting for us in the full NNLO calculation. This complexity
does in fact reveal itself in terms of a very rich polylogarithmic structure of the 
Laurent series expansion of the scalar one-loop integrals as well as the appearance
of multiple polylogarithms of maximal weight and depth four.

To underscore the statement that the loop--by--loop contributions reveal the
mathematical structure of a full NNLO calculation let us take a look at the paper by
Bernreuther {\it et al.} \cite{bernreuther04} who calculated the 
${\cal O}(\epsilon^1)$ contributions to the one--loop vertex correction of the
process $V \to Q \bar{Q}$. These are needed for the loop--by-loop part of a NNLO
calculation of heavy quark pair production in $e^+-e^-$--annihilations. The result 
can be expressed in terms of one--dimensional harmonic polylogarithms of maximal 
weight three. The same paper also lists results on the corresponding 
${\cal O}(\epsilon^0)$ two--loop 
vertex corrections which contain one--dimensional harmonic polylogarithms of maximal 
weight four. This shows that the same mathematical complexity appears in the
loop--by--loop contributions as in the two--loop contribution. Let me mention that 
heavy quark pair production in $e^+-e^-$--annihilations 
is a somewhat simpler problem than heavy quark pair production in hadronic collisions 
because of the appearance of one additional mass scale in the latter case. 
This explains 
why one has only one--dimensional harmonic polylogarithms in $e^+-e^-$--annihilations 
case compared to the multiple polylogarithms appearing in the hadronic collision 
calculation. Even then the two--loop vertex correction to $V \to Q \bar{Q}$ 
listed in \cite{bernreuther04} takes up more than twelve pages.   

The scalar four--point integrals appearing in the calculation of the loop--by--loop
evaluation are the most difficult to calculate. They contain a
very rich structure in terms of polylogarithmic functions. For example, the 
$\epsilon^2$--coefficients of the Laurent series expansion of the
four--point integrals contain logarithms and classical 
polylogarithms up to order four ({\it i.e.} $ Li_4 $) in conjunction with 
the $\zeta$--functions $\zeta(2),\zeta(3)$ and $\zeta(4)$ and products thereof, 
and a new class of 
functions which are now termed multiple polylogarithms \cite{gon}.

Since this is a conference
on mathematical physics it is appropiate to dwell a little
on the subject of multiple polylogarithms. A multiple polylogarithm is represented by
\begin{eqnarray}
\lefteqn{Li_{m_{k},...,m_{1}}(x_{k},...,x_{1})=
\int \limits_{0}^{x_{1}x_{2}...x_{k}} \left( \frac{dt}{t} \circ \right)^{m_{1}-1}} \nonumber \\ 
& &\frac{dt}{x_{2}x_{3}...x_{k}-t} \circ 
\left( \frac{dt}{t} \circ \right)^{m_{2}-1} 
                                                                     \nonumber \\ 
& &  \frac{dt}{x_{3}...x_{k}-t} 
\circ ...  \circ  \left( \frac{dt}{t} \circ \right)^{m_{k}-1}   \frac{dt}{1-t}\; ,
\nonumber
\end{eqnarray}
where the iterated integrals are defined by
\begin{eqnarray}
\lefteqn{\int \limits_{0}^{\lambda} \frac{dt}{a_{n}-t}\circ ...\circ  
\frac{dt}{a_{1}-t}=} \nonumber \\
& & \!\!\!\!\!\!\int \limits_{0}^{\lambda} \frac{dt_{n}}{a_{n}-t_{n}}
\int \limits_{0}^{t_{n}} \frac{dt_{n}}{a_{n-1}-t_{n-1}} \times...\times \int \limits_{0}^{t_{2}}\frac{dt}{a_{1}-t_{1}} \;.\nonumber
\end{eqnarray}
The indices $m_k$ and $k$ are positive integers. The multiple polylogarithms are 
classified according to their weight $w= m_1+m_2+ ... +m_k$ and their depth $k$.
We mention that a very efficient program for the numerical evaluation of multiple
polylogarithms has recently been developed in Mainz which, characteristically, is based 
on the language GiNaC \cite{vw05}.
  
The classical polylogarithms, Nielsen's generalized polylogarithms, the one-- and 
two--dimensional harmonic polylogarithms are all special cases of Goncharov's
multiple polylogarithms (see e.g. \cite{rogal05}). For example, the classical polylogarithms
\begin{equation}
Li_n(z)=\int_0^z\frac{Li_{n-1}(x)}{x} dx \hspace{0.5cm} n \ge 2 \,\,; 
\hspace{0.5cm} Li_1(z)=-\ln(1-z)
\end{equation}
are multiple polylogarithms of weight $n$ and depth 1.

In our original Feynman parameter calculation our results were written down as 
one--dimensional integral representations given by the integrals 
\begin{equation}
F_{\sigma_1\sigma_2\sigma_3}(\alpha_1,\alpha_2,\alpha_3,\alpha_4)=
\int_0^1 dy \frac{\ln (\alpha_1+\sigma_1 y) \ln (\alpha_2+\sigma_2 y)
\ln (\alpha_3+\sigma_3 y)}{\alpha_4+y} 
\end{equation}
and
\begin{equation}
F_{\sigma_1}(\alpha_1,\alpha_2,\alpha_3,\alpha_4)= 
\int_0^1 dy \frac{\ln (\alpha_1+\sigma_1 y) {\rm Li}_2(\alpha_2+\alpha_3 y)
}{\alpha_4+y}\ ,
\end{equation}
\noindent where the $\sigma_i $ take values $\pm 1$ and the $\alpha_j$'s are 
combinations of the kinematical variables of the process. The numerical evaluation
of these one-dimensional integral representations are quite stable. The functions
$F_{\sigma_1\sigma_2\sigma_3}(\alpha_1,\alpha_2,\alpha_3,\alpha_4)$ and
$F_{\sigma_1}(\alpha_1,\alpha_2,\alpha_3,\alpha_4)$ are related to multiple
polylogarithms of maximal weight and depth four as shown in \cite{kmr06a}.

We are now in the process of computing the full loop--by--loop contributions
including the spin and colour algebra arising from squaring the full 
one--loop amplitudes as given in \cite{kmr06b}. A first result has been obtained for 
the Abelian case of photon--photon production of heavy quark pairs \cite{kmr06c}.

\section{Decays of unpolarized and polarized top quarks}

After this brief mathematical detour I return to the physics of top quark decays. In 
the SM the top quark decays almost 100\% to a $W^+$ and a bottom quark. Also, the top
quark decays so fast that it retains its initial polarization when it decays. 
I describe both unpolarized and polarized top quark decays. In the unpolarized case 
I analyze top quark decays
in cascade fashion as a two step process involving the decay 
$t \rightarrow b + W^+$ and $W^+ \rightarrow l^+ + \nu_l$ in the respective
rest frames of the top quark and the $W^+$--boson. In the polarized case
I perform the decay analysis in the rest system of the top quark.  
I discuss polar and azimuthal correlations involving the polarization
of the top quark and the momenta of the decay products in the 
decay $t(\uparrow) \to X_b + l^+ + \nu_l$.

The decay of a polarized top quark into a $W^+$--boson and a jet with
$b$--quantum numbers $t(\uparrow) \to X_b + l^+ + \nu_l$ is desribed by altogether 
eight invariant structure functions (see e.g. \cite{Manohar:1993qn,fgkm02}).

 \begin{eqnarray} 
 \label{tensor-expansion}
   H^{\mu \nu} & = & \Big( - g^{\mu \nu} \, H_1 + p_t^{\mu} p_t^{\nu} \, H_2 -
   i \epsilon^{\mu \nu \rho \sigma} p_{t,\rho} q_{\sigma} \, H_3 \Big) + 
   \nonumber \\ & - &
   (q \!\cdot\! s_t) \Big( - g^{\mu \nu} \, G_1 + p_t^{\mu} p_t^{\nu} \, G_2 -
   i \epsilon^{\mu \nu \rho \sigma} p_{t,\rho} q_{\sigma} \, G_3 \Big) +
   \\ & + &
   \Big(s_t^{\mu} p_t^{\nu} + s_t^{\nu} p_t^{\mu} \Big) \, G_6 +
   i \epsilon^{\mu \nu \rho \sigma} p_{t \rho} s_{t \sigma} \, G_8 +
   i \epsilon^{\mu \nu \rho \sigma} q_{\rho} s_{t \sigma} \, G_9 \, , \nonumber
 \end{eqnarray}
 There are three unpolarized
structure functions $H_{1,2,3}$ and five polarized 
structure functions from the set $G_{1,2,3,6,8,9}$ \footnote {In physical
expressions the three structure functions $G_3, G_8$ and $G_9$  
contribute only in two pairs of linear combinations \cite{fgkm02}}. 
In general the invariant structure 
functions are functions of $q_0$ and $q^2$. In the narrow resonance 
approximation for the $W^+$--boson, which we shall adopt in this talk,
one has $q^2=m_W^2$. The aim of the game is to measure the different
unpolarized and polarized structure functions (or moments thereof) and to
compare them to theoretical predictions. The different structure functions can be 
separated since they contribute to the rate with
different dependencies on the electron energy and, in the 
case of the polarized structure functions, they can be measured through polar 
and azimuthal 
correlations involving the polarization direction of the polarized top 
quark.

\section*{Unpolarized top quark decays}

\begin{figure}[!h]
\begin{center}
\epsfig{figure=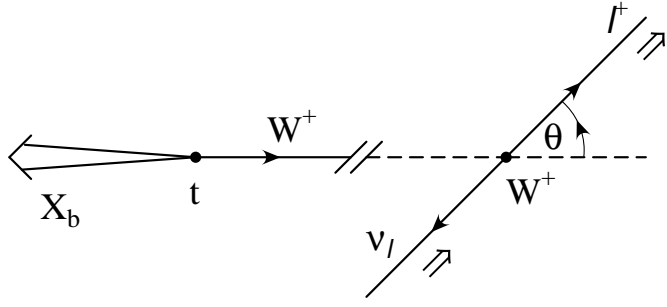, scale=0.5}
\caption{Definition of the polar angle $\theta$ in the rest frame decay of
$W^+ \rightarrow l^+ + \nu_l$. The two lines ``//'' indicate a boost to the rest
system of the $W^+$. The arrows next to the lepton lines give the helicities of the
leptons.}
\label{theta}
\end{center}
\end{figure}

In the decay $t \to X_b + W^+$ the $W^+$ is polarized. The $W^+$ is 
self--analyzing in the sense that the angular decay
distribution of its decay products $W^+ \to l^+\,\nu_l$ 
can be used to reconstruct the polarization of the $W^+$.
We shall analyze the unpolarized decay in cascade--type fashion, i.e.
we shall analyze the decay $W^+ \rightarrow l^+ + \nu_l$ in the rest frame of
the $W^+$. This brings in the three unpolarized helicity structure functions 
$H_{T_+},H_{T_-},H_L$ (or for short $H_+,H_-,H_L$) which are linearly related to 
the three unpolarized invariant
structure functions $H_1, H_2, H_3$ via \cite{fgkm02}
\begin{eqnarray}
 \label{helicity}
   H_+ & = &  H_1 + |\vec{q} \,| m_t \, H_3 , \\  
   H_- & = &  H_1 - |\vec{q} \,| m_t \, H_3 , \\
   H_L & = & m_W^2 H_1 + |\vec{q} \,|^2 m_t^2 \, H_2,  
 \end{eqnarray}
The polar angle decay distribution is given by 
\vspace{-0.1cm}  
\begin{equation}
\label{polar}
\frac{1}{\Gamma} \frac{d \Gamma}{d \cos \theta} = \frac{3}{8} 
(1+ \cos \theta)^2 \; {\cal H}_+ + \frac{3}{8}(1- \cos \theta)^2 \; {\cal H}_-
 + \frac{3}{4} \sin^2 \theta \; {\cal H}_L \; . \nonumber
\end{equation}
where the angle $\theta$ is defined in Fig.\ref{theta}. The ${\cal H}_+, {\cal H}_-$ 
and ${\cal H}_L$ are the normalized transverse--plus, 
transverse--minus and longitudinal helicity structure functions, resp.,
such that ${\cal H}_+ +{\cal H}_- + {\cal H}_L =1$.

From the polar angle dependence in Eq.(\ref{polar}) or from matching $m$--quantum 
numbers in the $W^+$ rest frame decay (see Fig.\ref{theta}) it is clear that
\begin{eqnarray}
{\cal H}_+ &:& \quad \mbox{favours forward}\quad l^+ 
\nonumber \\
{\cal H}_- &:& \quad \mbox{favours backward}\quad l^+ 
\nonumber
\end{eqnarray} 
Translated to the top quark rest frame this implies that ${\cal F}_+$ (${\cal F}_-$)
produce harder (softer) $l^+$'s which can be used to experimentally separate the
contributions of the three helicity structure functions.

At the Born term level the SM prediction is ($m_t=175 \, \mbox{GeV}$, $m_b = 0$)

\begin{eqnarray}
\qquad \qquad \qquad {\cal H}_+(\mbox{Born})  & = & 0  \qquad \quad \mbox{(forbidden)} \\
\qquad \qquad \qquad {\cal H}_-(\mbox{Born})  & = & \frac{1}{1+2y^2}\;=\;0.297 
\nonumber \\
\qquad \qquad \qquad {\cal H}_L(\mbox{Born})  & = & \frac{2y^2}{1+2y^2}\;=\;0.703 \, , \nonumber
\end{eqnarray}
where $y=m_W/m_t$. At the Born term level, with 
$m_b = 0$, ${\cal H}_+$ is not populated because of angular
momentum conservation in the two--body decay process $t \to b + W^+$ where for
$m_b = 0$ the bottom quark has 100\% negative helicity and the 
$b$--quark and the $W^+$ are in a back--to--back configuration. 

The present experimental results on ${\cal H}_+$ are consistent with zero within 
large error bars. For example, using 230 ${\mbox{pb}}^{-1}$ D0 quotes a value of
${\cal H}_+= 0.00 \pm 0.13\,({\mbox{stat}}) \pm 0.07 \,({\mbox{syst}})$ 
\cite{abazov05a}. 
CDF finds ${\cal H}_+= 0.00 {+0.22\atop -0.34}({\mbox{stat + syst}})$ or 
${\cal H}_+ < 0.27$ at the 95\% confidence level \cite{abulencia06}. Using 
the same data sample of
200 ${\mbox{pb}}^{-1}$ CDF quotes  a value of ${\cal H}_L = 0.74 {+0.22\atop -0.34}$
for the longitudinal helicity of the $W^+$--boson, also compatible with the SM
prediction.

The vanishing of ${\cal H}_+$ is no longer 
true for additional gluon or photon emission, or when one takes into account 
bottom mass effects. When all of these are taken into account one has 
\cite{fgkm01,dgkm03}

\begin{equation}
{\cal H}_+ = 0.00102 (\mbox{QCD}) + 0.00008(\mbox{EW}) + 0.00039(m_b\neq 0),
\end{equation}
where the numbers give the ${\cal O}(\alpha_s)$ \,  QCD corrections, the
${\cal O}(\alpha)$ electroweak corrections and $m_b \neq 0$ corrections
($m_b=4.8 \, {\mbox {GeV}}$). 
Numerically the correction to ${\cal H}_+$ occurs only at the pro mille
level. It is safe to say that, if top quark decays reveal a violation of the SM
$ (V\!-\!A) $ current structure that exceeds the $ 1\% $ level, the violations
must have a non-SM origin. 

The results for the corresponding corrections to ${\cal H}_-$ and ${\cal H}_L$ are 
listed in terms of rates normalized to the total Born term
rate, i.e. $\hat{\Gamma}_i=\Gamma_i / \Gamma(\mbox{Born})$. The normalized 
partial Born term rates $\hat{\Gamma}_i(\mbox{Born})$
are factored out. Corrections coming from NLO QCD, from the NLO electro-weak
corrections (EW), from the $W^+$ finite width correction (BW) and
from $m_b \neq 0$ effects are listed separately. One has
\vspace{-0.5cm}
 
\begin{eqnarray}
\hat{\Gamma}_- \!&=&\! 0.297\left[1 - 0.0656(\mbox{QCD}) + 0.0206(\mbox{EW})
- 0.0197(\mbox{BW}) - 0.00172(m_b \neq 0)\right] \nonumber \\
\hat{\Gamma}_L \! &=&\! 0.703\left[1 - 0.0951(\mbox{QCD}) + 0.0132(\mbox{EW})
- 0.0138(\mbox{BW}) - 0.00357(m_b \neq 0)\right] \nonumber
\end{eqnarray}
Written in terms of the normalized ${\cal H}_i$ this translates into a 
\,+2.4\% upward 
shift from ${\cal H}_-(\mbox{Born})=0.297$ and a \,-1.2\% downward shift from
${\cal H}_L(\mbox{Born})=0.703$. Judging from the fact that ${\cal H}_L$ and
${\cal H}_-$ will eventually be measured to better than 1\% it is quite
clear that one has to take radiative corrections into account when comparing experiment 
with theory.
   
\begin{figure}[!h]
\begin{center}
\epsfig{figure=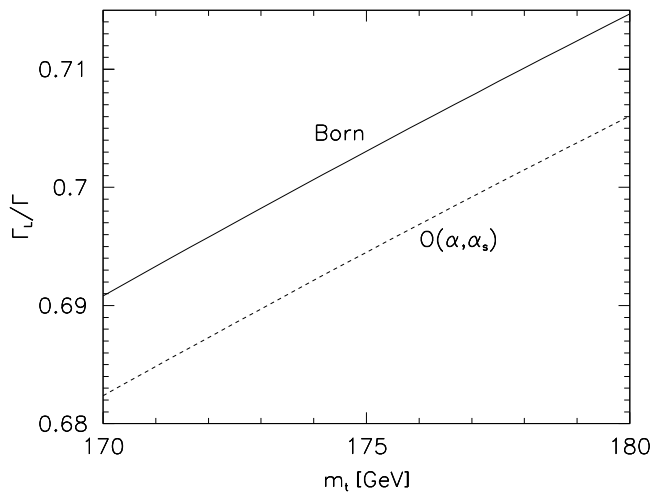, scale=0.5}
\caption{Top mass dependences of the ratio
   ${\cal H}_L= \Gamma_{L}/\Gamma $. Full line : Born term.
   Dashed line: Corrections including (QCD),
  electroweak (EW), finite-width (FW) and ($ m_b \ne 0 $) Born term
   corrections.}
\label{topmassdep}
\end{center}
\end{figure}

In Fig.\ref{topmassdep} we show the top mass dependence of the ratio 
${\cal H}_L =\Gamma_L/\Gamma$.
 The horizontal displacement of the Born term curve and the corrected curve
 is $ \approx 3.5 $ GeV. One would thus make the corresponding mistake
 in a top mass determination from the measurement of ${\cal H}_L$ 
 if the Born term curve was used instead of the corrected curve.
 If one takes $ m_t=175 $ GeV as central value a $ 1\% $ relative error on
 ${\cal H}_L$ would allow one to determine the top
 quark mass with an error of $ \approx $ 3 GeV.
\section{Polarized top quark decays}

Contrary to the analysis of unpolarized top quark decays described in the last
subsection polarized top quark decay will be analyzed altogether in the rest frame 
of the decaying top quark. This is the natural choice for an experimental
analysis. Choosing a particular two--particle rest subsystem is only of advantage
if that particular subsystem is resonance dominated as was discussed in the 
unpolarized decay case.

\begin{figure}[t]
\begin{center}
\epsfig{figure=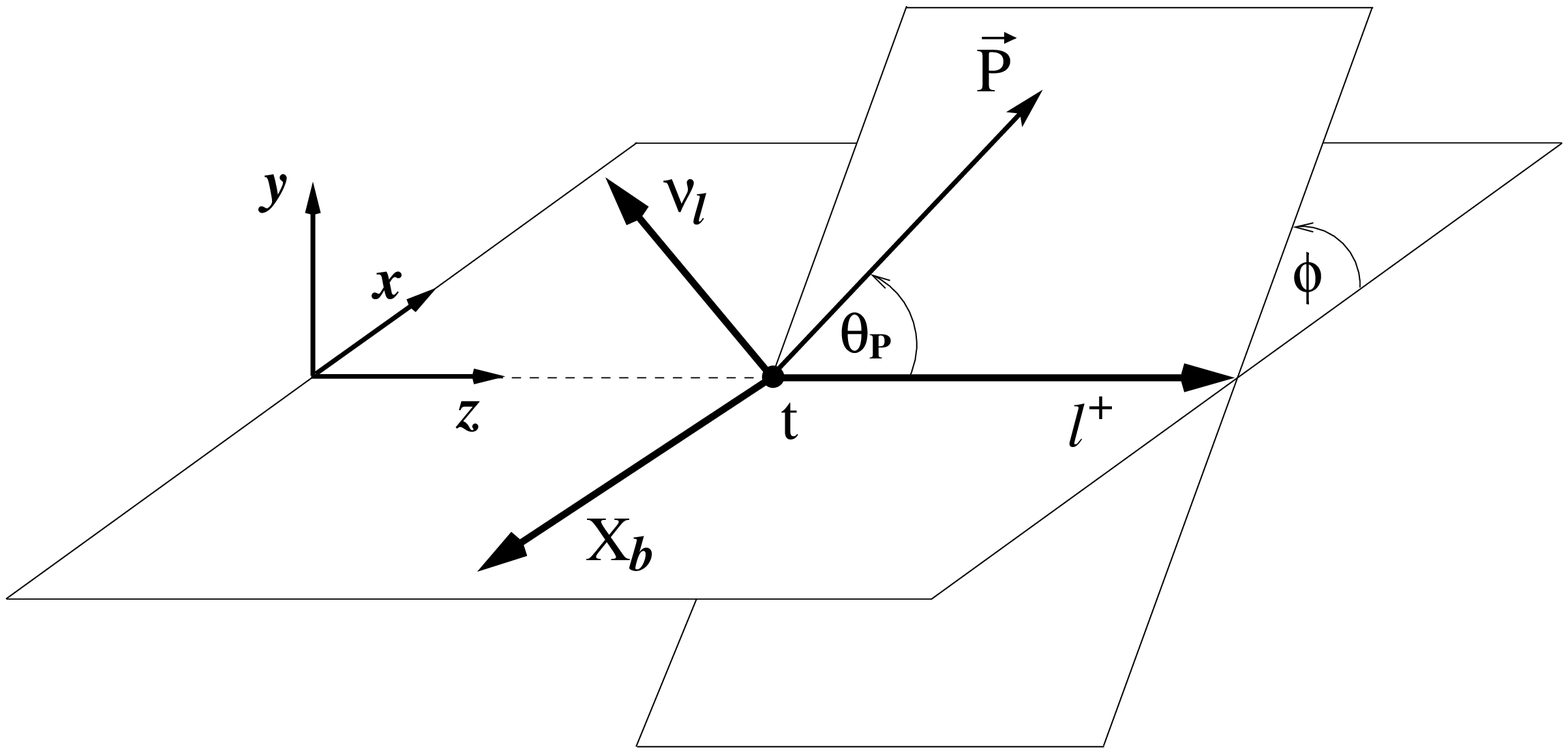, scale=0.3}
\vspace{7pt}\centerline{(1a)}\vspace{7pt}
\epsfig{figure=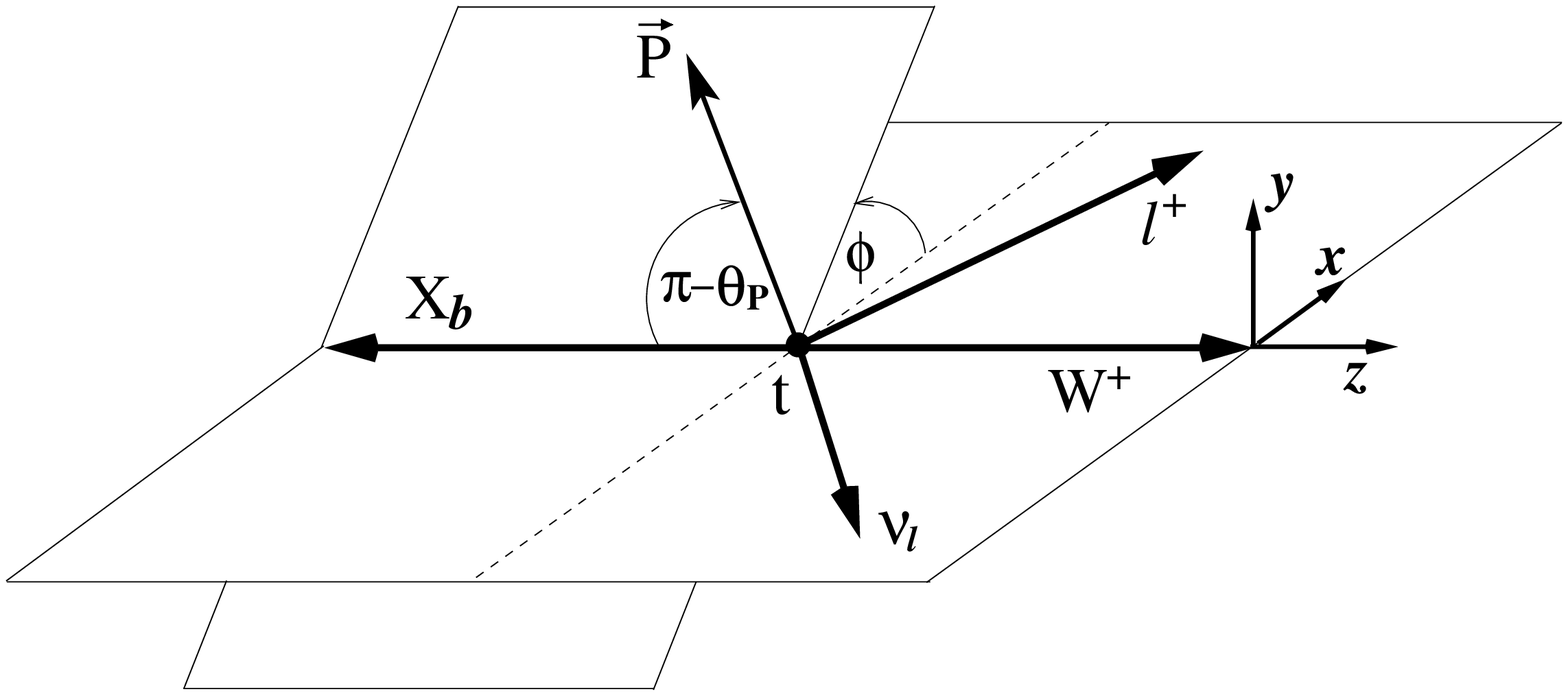, scale=0.3}
\vspace{7pt}\centerline{(2'a)}\vspace{7pt}
\epsfig{figure=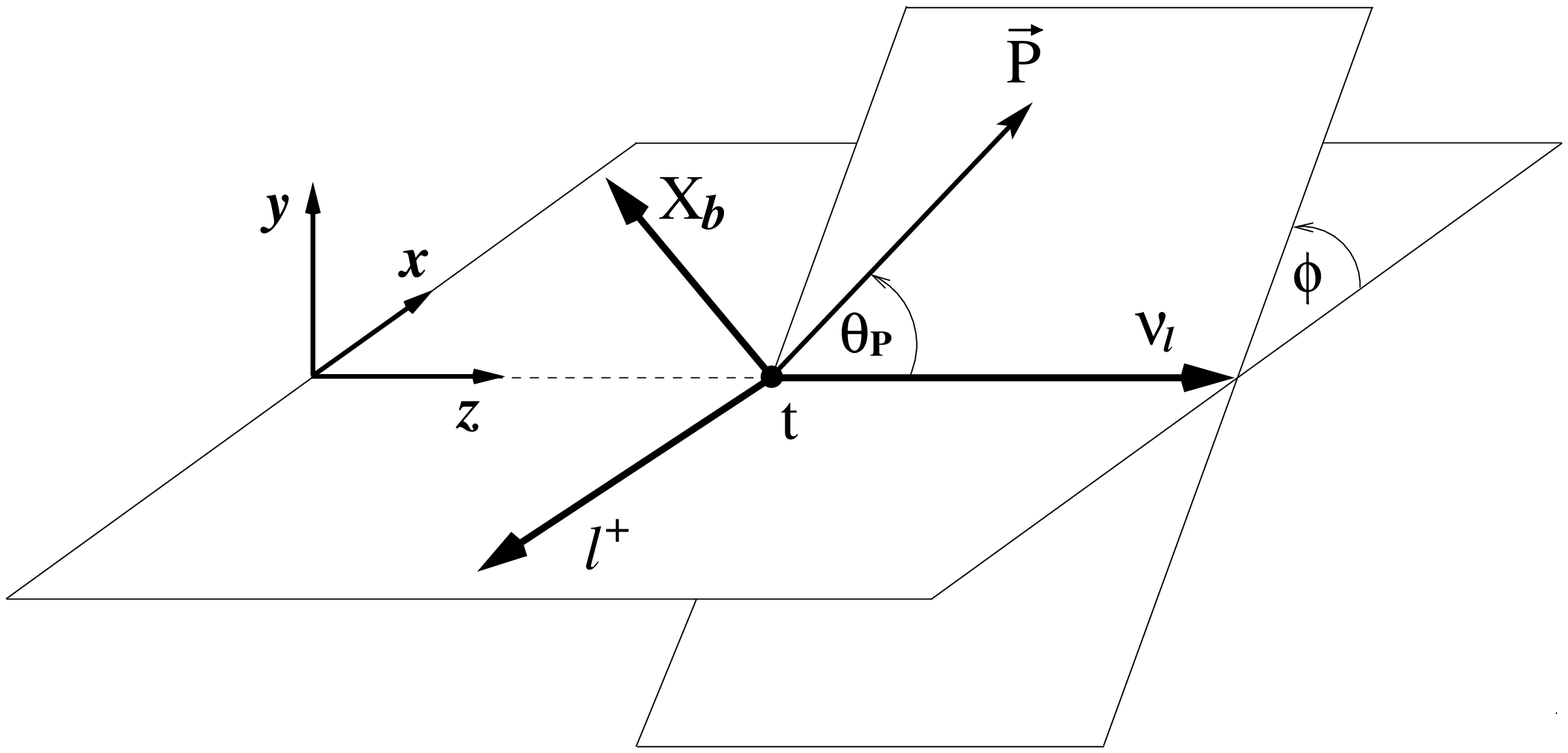, scale=0.3}
\vspace{7pt}\centerline{(3a)}\vspace{7pt}
\caption{The definition of the polar angle $\theta_P$ and the azimuthal 
angle $\phi$ in the rest frame decay of a polarized top quark in three different
helicity systems. 
The event plane defines the $(x,z)$--plane with (1a) $ \vec{p}_l \parallel z $
and $(\vec{p}_\nu)_x \ge 0$, $(2^\prime a$) $ \vec{q} \parallel z $ and 
$(\vec{p}_l)_x \ge 0$, and (3a) $ \vec{p}_\nu \parallel z $ 
and $(\vec{p}_l)_x \le 0$.}
\label{helframes}
\end{center}
\end{figure}

The general angular decay distribution of the rest frame decay of a polarized 
top quark decaying into a jet $X_b$ and a lepton $l^+$ and a neutrino is given 
by \cite{kp99}

\begin{equation}
\frac{d\Gamma}{dx_l d\hat{q}_0d\cos\theta_Pd\phi}= 
\frac{1}{4\pi}\left(\frac{d\Gamma_A}{dx_ld\hat{q}_0}
+ P(\frac{d\Gamma_B}{dx_ld\hat{q}_0}\cos\theta_P
+\frac{d\Gamma_C}{dx_ld\hat{q}_0}\sin\theta_P \cos\phi) \right) \end{equation}

\noindent where the polar and azimuthal angles $\theta_P$ and $\phi$ describe the 
orientation of the polarization of the top quark relative to the decay plane 
formed by the decay products of the top quark.
The scaled energy and the scaled mass of the 
$W^+$ are denoted by $\hat{q}_0=q_0/m_t$ and $y=m_W/m_t$.  As usual we define a 
scaled lepton energy
through $x_l=2E_l/m_t$. $P$ is the magnitude of the 
top quark polarization. $\Gamma_A$ stands for the unpolarized rate, and $\Gamma_B$
and $\Gamma_C$ stand for the polar and azimuthal correlation rates. In \cite{ghkkk06} 
we have considered three different
helicity systems to analyse the polar and azimuthal correlations in the rest 
frame decay of a polarized top quark as shown in Fig.\ref{helframes}.

It is important to realize that correlation measurements in each of the helicity frames
constitute independent measurements of the invariant polarized structure 
functions. To illustrate this point let us consider the contribution of the 
invariant polarized structure
function $G_1$ to the polar and azimuthal correlations in the above three 
helicity systems. The decay rate is proportional to 
$L^{\mu \nu}H_{\mu \nu}$. One
then obtains
\begin{equation}
\label{illustration}
L^{\mu \nu}H_{\mu \nu}(G_1)= m_t q^2 G_1 
\left\{ 
\begin{array}{l} 
{\displaystyle \frac{x_l\hat{q}_0-y^2}{x_l}} \cos \theta_{P1} + 
{\displaystyle \frac{y}{x_l}}
\sqrt{x_l(2\hat{q}_0-x_l)-y^2} \sin\theta_{P1}\cos\phi \vspace*{2mm} \\
 \sqrt{\hat{q}_0^2-y^2}\cos \theta_{P2}  \\
{\displaystyle \frac{\hat{q}_0-y^2}{2\hat{q}_0-x_l}} \cos \theta_{P3}
+ {\displaystyle \frac{y}{2\hat{q}_0-x_l}} \sqrt{x_l(2\hat{q}_0-x_l)-y^2}
\sin\theta_{P3}\cos\phi 
\end{array} 
\right\}\, ,
\end{equation}
where the three contributions in the curly bracket refer to the polar
and azimuthal correlations in the three helicity coordinate systems 
with the $z$--axes along (1) the lepton $l^+$\,, (2) the $W^+$--boson
\, and (3) the neutrino $\nu_l$\,. From Eq.(\ref{illustration}) it is clear
that $G_1$ contributes quite differently to the correlation functions in the
three reference systems. 

In \cite{ghkkk06} we have calculated the Born term and NLO QCD contributions
to the polar and azimuthal correlation functions $d \Gamma_B$ and $d \Gamma_C$
in the three different helicity systems. We were able to obtain closed form expressions
for the totally integrated angular decay distributions. The results are too long to be 
listed here but can be found in \cite{ghkkk06}. We mention that we find agreement
with \cite{jk89,cjk91,cjkk94,cj94} for the unpolarized case
$d \Gamma_A$ and the polar correlation function $d \Gamma_B$ in systems 1 and 3. In numerical form one
has   \\
\underline{$z$--axis along lepton} (system (1a)) \\
\begin{equation}
\label{numbersIa}
\frac{d\Gamma^{{\rm NLO}}}{d \cos\theta_P \, d\phi}=
\frac{\Gamma_A^{(0)}}{4\pi}\bigg[(1-8.54\%)+ (1-8.72\%) P \cos\theta_P - 0.24\%
\,P \sin\theta_P \cos\phi \bigg]
\end{equation} 
\underline{$z$--axis along $W^+$--boson} (system ($2^\prime a$))
\begin{equation}
\label{numbersIIa}
\frac{d\Gamma^{{\rm NLO}}}{d \cos\theta_P \, d\phi}=
\frac{\Gamma_A^{(0)}}{4\pi}\bigg[(1-8.54\%)+ (0.406-11.62\%) P \cos\theta_P - 
(0.760-8.20\%)
\,P \sin\theta_P \cos\phi \bigg]
\end{equation}
\underline{$z$--axis along neutrino} (system (3a))
\begin{equation}
\label{numbers3a}
\frac{d\Gamma^{{\rm NLO}}}{d \cos\theta_P \, d\phi}=
\frac{\Gamma_A^{(0)}}{4\pi}\bigg[(1-8.54\%)- (0.318-1.02\%) P \cos\theta_P - 
(0.919-8.61\%)
\,P \sin\theta_P \cos\phi \bigg]
\end{equation}
In all the three expressions we have factored out the Born term rate $\Gamma_A^{(0)}$.
The first number in the round brackets stands for the LO Born term rate whereas
the second number gives the percentage change due to the NLO \, QCD corrections.

Let me first discuss the LO correlation functions. I shall refer to $\Gamma_B/\Gamma_A$
and $\Gamma_C/\Gamma_A$ as the polar and azimuthal analyzing power, respectively. 
In system (1a) ($l^+$ along $z$)
the polar analyzing power is 100\% which necessarily implies that the azimuthal 
analyzing power is zero in this system. In fact, the vanishing of $\Gamma_C$ in system
(1a) can be seen to directly follow from the left--chiral $(V-A)$ structure of the SM 
quark and lepton currents \cite{ghkk06}. The polar analyzing power in the systems
($2^\prime a$) and (3a) is less than 100\% with +41\% and -32\%, respectively. As 
mentioned before the
LO azimuthal analyzing power in system (1a) is zero. In system ($2^\prime a$) and 
(3a) the azimuthal analyzing power is reasonably large with -76\% and -92\%, 
respectively.    

Except for the polar correlation in system (3a) all NLO corrections go in the same 
direction. They reduce the LO results by approximately 10\%. This implies that
the polar and azimuthal analyzing powers are not changed very much from their Born
term values through  
radiative correction. An exception is system (3a) where the polar analyzing power
is changed from -31.8\% to -34.4\%. This amounts to a 8.2\% change in analyzing
power through radiative corrections which is surprisingly large.


\section{Summary and conclusions}

In this talk I have covered three selected topics in top quark physics. The first
topic concerned the NNLO calculation of hadronic top quark pair production where
the loop--by--loop part is now being completed. The other three missing
parts of the NNLO calculation (two--loop, one-loop gluon emission, two--gluon 
emission) are more difficult and will very likely take another five to 
ten years to complete. Such a large--size calculation will require a dedicated 
international effort of the theoretical
community which will have to be coordinated by one of the big international centers 
of particle physics. In the second and third topic I discussed NLO\, QCD 
predictions
for unpolarized and polarized top quark decays which should be amenable to 
experimentals tests in the next coming few years.

\vspace{1cm} {\bf Acknowledgements:}
I would like to thank my collaborators S.~Groote, W.~S.~Huo, A.~Kadeer, D.~Kubistin,
Z.~Merebashvili and M.~Rogal for participating in the work I have reported on
in this talk. My thanks are also due to Riazuddin and F. Hussain for organizing
such a wonderful conference in Islamabad, and to all the Pakistani graduate student
helpers who made the meeting such a joyful affair.


\begin{thebibliography}{11}
\bibitem{Dawson:1988}
P.~Nason, S.~Dawson and R.~K.~Ellis,
Nucl.\ Phys.\ {\bf B303} (1988) 607; {\it ibid} {\bf B327} (1989) 49;
{\it ibid} {\bf B335} (1990) 260(E).
\bibitem{been}
W.~Beenakker, H.~Kuijf, W.~L.~van Neerven and J.~Smith, Phys. Rev. D
{\bf 40} (1989) 54 ; W.~Beenakker, W.~L.~van Neerven, R.~Meng, G.A.~Schuler
and J.~Smith, Nucl. Phys. {\bf B351} (1991) 507. 
\bibitem{LO:1978}
M.~Gl\"uck, J.F.~Owens and E.~Reya, Phys. Rev. D {\bf 17} (1978) 2324; \\
B.~L.~Combridge, Nucl.\ Phys.\ {\bf B151} (1979) 429; \\
J.~Babcock, D.~Sivers and S.~Wolfram, Phys. Rev. D {\bf 18} (1978) 162 ; \\
K.~Hagiwara and T.~Yoshino, Phys. Lett. {\bf 80B}, (1979) 282; \\
L.~M.~Jones and H.~Wyld, Phys. Rev. D {\bf 17} (1978) 782; \\
H.~Georgi {\it et al.}, Ann. Phys. (N.Y.) {\bf 114} (1978) 273 .
\bibitem{Italians}
M.~Cacciari, S.~Frixione, M.~L.~Mangano, P.~Nason, G.~Ridolfi, 
JHEP {\bf 0407} (2004) 033.
\bibitem{acosta04} D.~Acosta {\it et al.}, The CDF Collaboration,  Phys.~Rev.~Lett.
{\bf 93} (2004) 142001.
\bibitem{abazov05} V.~M.~Abazov {\it et al.}, The D0 Collaboration, 
Phys. Lett. {\bf 626B}, (2005) 55
\bibitem{cacciari04}
  M.~Cacciari, S.~Frixione, M.~L.~Mangano, P.~Nason and G.~Ridolfi,
  JHEP {\bf 404} (2004) 068
\bibitem{kidonakis03}
N.~Kidonakis and R.~Vogt,
Phys. Rev. D {\bf 68} (2003) 114014.
\bibitem{km02}
J.G. K\"orner and Z. Merebashvili, Phys. Rev. D {\bf 66} (2002) 054023.
\bibitem{kmr1}
J.G. K\"orner, Z. Merebashvili and M. Rogal,  
Phys.\ Rev.\ D {\bf 71} (2005) 054028.
\bibitem{kmr06a}
  J.~G.~K\"orner, Z.~Merebashvili and M.~Rogal,
J.~Math.~Phys. {\bf 47} (2006) 072302,  arXiv:hep-ph/0512159.
\bibitem{bernreuther04}
  W.~Bernreuther, R.~Bonciani, T.~Gehrmann, R.~Heinesch, T.~Leineweber, P.~Mastrolia and E.~Remiddi,
  Nucl.\ Phys.\ B {\bf 706} (2005) 245
\bibitem{gon}
A.B.~Goncharov, Math. Res. Lett. {\bf 5} (1998), available at
http://www.math.uiuc.edu/K-theory/0297 .
\bibitem{vw05}
J.~Vollinga and S.~Weinzierl,
  Comput.\ Phys.\ Commun.\  {\bf 167} (2005) 177, arXiv:hep-ph/0410259.
\bibitem{rogal05}
M.~Rogal, doctoral thesis, Mainz 2005. Available at  
http://wwwthep.physik.uni-mainz.de \,.
\bibitem{kmr06b}
J.G. K\"orner, Z. Merebashvili and M. Rogal, 
  Phys.\ Rev.\ D {\bf 73} (2006) 034030.
\bibitem{kmr06c}
J.G. K\"orner, Z. Merebashvili and M. Rogal, hep-ph/0608287
\bibitem{Manohar:1993qn}
  A.~V.~Manohar and M.~B.~Wise, Phys.\ Rev.\ D {\bf 49} (1994) 1310
\bibitem{fgkm02}
  M.~Fischer, S.~Groote, J.~G.~K\"orner and M.~C.~Mauser,
  Phys.\ Rev.\ D {\bf 65} (2002) 054036
\bibitem{abazov05a} V.~M.~Abazov {\it et al.}, The D0 Collaboration, 
Phys.~Rev.~D {\bf 72}, (2005) 011104.
\bibitem{abulencia06}
  A.~Abulencia {\it et al.}  The CDF-Run II Collaboration,
  Phys.\ Rev.\ D {\bf 73} (2006) 111103
\bibitem{fgkm01}
  M.~Fischer, S.~Groote, J.~G.~K\"orner and M.~C.~Mauser,
  Phys.\ Rev.\ D {\bf 63} (2001) 031501
  [arXiv:hep-ph/0011075].
\bibitem{dgkm03}
  H.~S.~Do, S.~Groote, J.~G.~K\"orner and M.~C.~Mauser,
  Phys.\ Rev.\ D {\bf 67} (2003) 091501
  [arXiv:hep-ph/0209185].
\bibitem{kp99} 
J.~G.~K\"orner and D.~Pirjol, Phys.\ Rev.\ 
D {\bf 60} (1999) 014021.

\bibitem{ghkkk06}
  S.~Groote, W.~S.~Huo, A.~Kadeer J.~G.~K\"orner and D.~Kubistin, to be published

\bibitem{ghkk06}
  S.~Groote, W.~S.~Huo, A.~Kadeer and J.~G.~K\"orner,
  arXiv:hep-ph/0602026.

\bibitem{jk89} M.~Jezabek and J.~H.~K\"uhn, Nucl. Phys. B {\bf 314} (1989) 1.

\bibitem{cjk91} A.~Czarnecki, M.~Jezabek and J.~H.~K\"uhn,
Nucl. Phys. B {\bf 351} (1991) 70.  

\bibitem{cjkk94} A.~Czarnecki, M.~Jezabek, J.~G.~K\"orner and J.~H.~K\"uhn, 
Phys. Rev. Lett. {\bf 73} (1994) 384.

\bibitem{cj94} A.~Czarnecki and M.~Jezabek,  Nucl. Phys. B {\bf 427} (1994) 3.

\end{thebibliography}
\end{document}